\documentclass[nato]{crckbked}

\usepackage{klucite}
\usepackage[dvips]{graphicx}

\def\idline{} 

\numreferences

\begin{document}

\begin{article}

\begin{opening}
\title{On the Multi-channel Anderson Impurity Model of Uranium Compounds}

\author{N.    \surname{Andrei}}
\author{C. J. \surname{Bolech}}
\institute{Rutgers University}

\runningauthor{N. Andrei, C. J. Bolech}

\begin{abstract}
In this talk we will present the solution of the two-channel Anderson
impurity model, proposed in the context of the heavy fermion compound
UBe$_{13}$, and discuss briefly the more general multi-channel
case. We will show results for the thermodynamics in the full range of
temperature and fields and make the connection with the current
experimental situation.
\end{abstract}
\end{opening}


\section{Models of Uranium ions}

During the last two decades, there has been growing interest in
 materials whose desciption falls outside  the framework of Landau's
Fermi Liquid Theory. Examples are the high T$_c$ superconductors, heavy
fermions and quasi-1D conductors.

Uranium heavy fermion compounds provide very interesting examples,
exhibiting unusual specific heat temperature dependence, $\gamma = C_V / T 
 \sim \ln T $. This behavior is often related to multichannel Kondo physics,
occasioned in the case of Uranium by the ground state configuration which
is degenerate and non-magnetic, so that one is lead to consider a
quadrupolar Kondo scenario \cite{cox87,ramirez94}.

\begin{figure}[tbh]
\label{hund1}
\begin{center}
\includegraphics[width=2.4in]{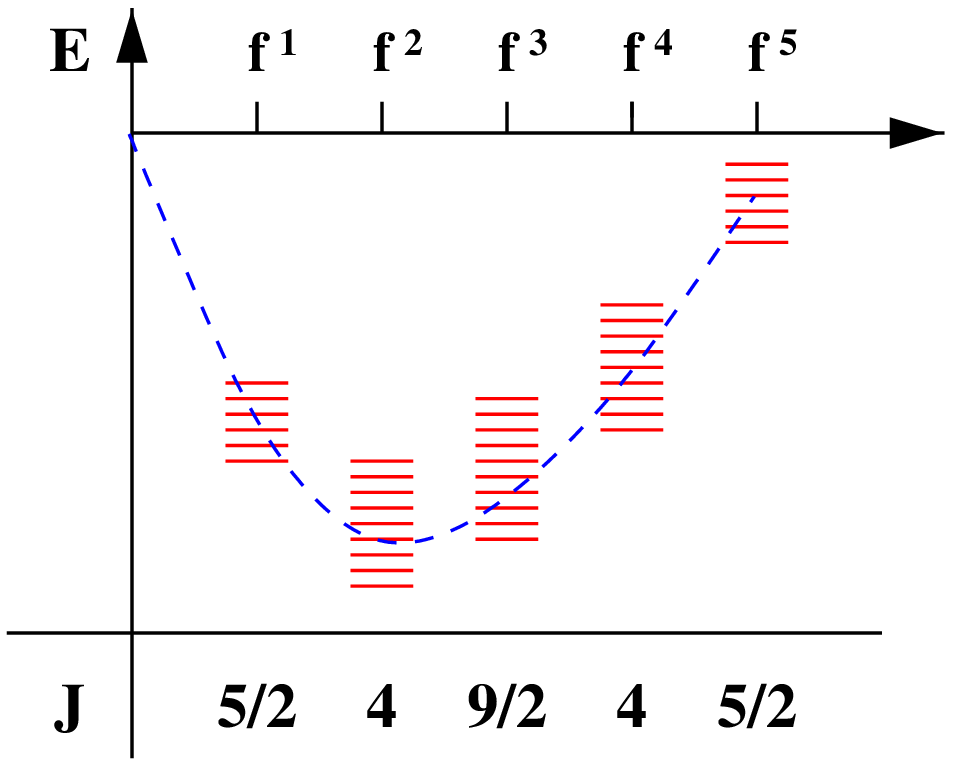}
\caption{Schematics of the energy and lowest angular momentum multiplet for
different states of valence of a uranium ion.}
\end{center}
\end{figure}

\begin{figure}[tbh]
\label{hund2}
\begin{center}
\includegraphics[width=2.7227in]{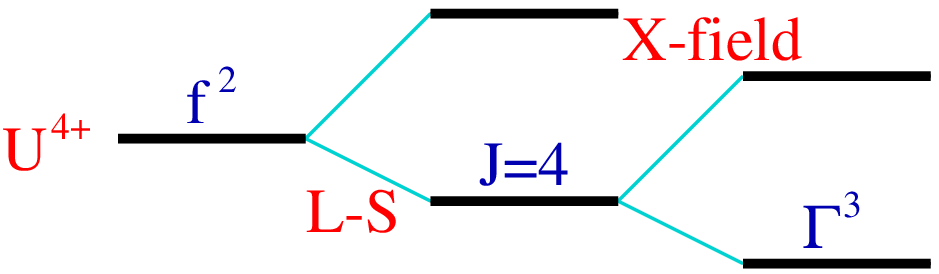}
\caption{Schematic depiction of the spin-orbit and crystal field splittings
for the ground state valence configuration of uranium in a cubic medium.}
\end{center}
\end{figure}

The starting point is to consider the energies of the uranium ion in
its different states of valence. They fall on a parabola 
depicted in Fig.\thinspace\ref{hund1}. The ground state corresponds to
a $\rm{U}^{4+}$ ionization state, with a $5\rm{f}^{2}$
shell-configuration. Considering the spin-orbit Hund's coupling, the
lowest multiplet corresponds to $J=4$. Taking, further, into account the
splitting due to a cubic crystal field (as is the case for the heavy
fermion compound \rm{UBe}$_{13}$) one finds that the lowest multiplet
corresponds to a non-Kramers $\Gamma_{3}$ doublet (see Fig.\thinspace
\ref{hund2}).

\begin{figure}[tbh]
\label{H738}
\begin{center}
\includegraphics[width=4.0736in]{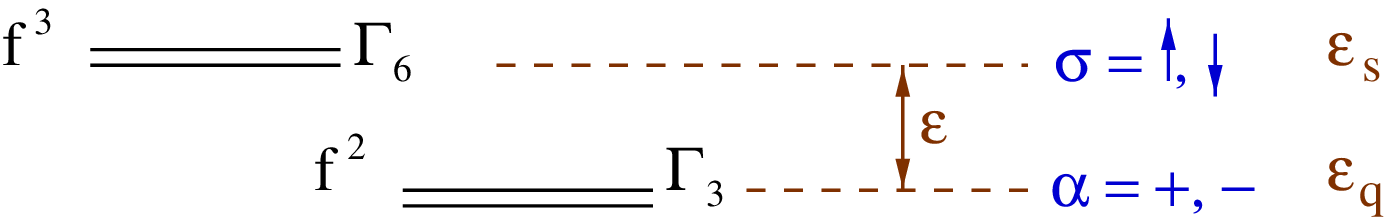}
\caption{Schematics of the electronic configurations and symmetry multiplets
retained for the Hamiltonian of \textrm{UBe}$_{13}$.}
\end{center}
\end{figure}

A similar analysis for the neighboring states of valence indicates
that they corresponds to: (\textbf{i}) $\rm{U}^{3+}$: a $5\rm{f}^{3}$
shell-configuration that splits giving a $J=9/2$ lowest multiplet
configuration, that further splits giving a lowest $\Gamma_{6}$
Kramer's doublet and (\textbf{ii}) $\rm{U}^{5+}$: a $5\rm{f}^{1}$
shell-configuration that splits giving a $J=5/2$ lowest multiplet
configuration, that further splits giving a lowest $\Gamma_{7}$
Kramer's doublet. These three valence states are in fact very close
in energy and their relative positions are not completely resolved,
since the experimental evidence is not conclusive and sometimes
contradictory \cite{aliev95,sac98}.  The most accepted scenario is
that of a $\Gamma_{3}$ quadrupolar doublet ground state coming from
the tetravalent state and next a $\Gamma_{6}$ magnetic doublet coming
from the trivalent state of uranium. Neglecting the contributions from
the pentavalent state of the uranium as well as those from excited
crystalline electric field states we can write down the following
Hamiltonian, the {\it two-channel Anderson impurity model} (see
Fig.\thinspace\ref{H738} for a schematic depiction).

\begin{eqnarray*}
H_{\Gamma_3,\Gamma_6} = H_{{\rm{host}}}&+&\varepsilon_{q}\sum_{\alpha
}b_{\bar{\alpha}}^{\dagger}b_{\bar{\alpha}}+\varepsilon_{s}\sum_{\sigma
}f_{\sigma}^{\dagger}f_{\sigma} \\
  &+&V\sum_{k,\alpha,\sigma}\left[  f_{\sigma
}^{\dagger}b_{\bar{\alpha}}c_{k,\alpha,\sigma}+c_{k,\alpha,\sigma}^{\dagger
}b_{\bar{\alpha}}^{\dagger}f_{\sigma}\right]  \\
\rm{with\qquad}Q &=& \sum_{\alpha}b_{\bar{\alpha}}^{\dagger}b_{\bar{\alpha}
}+\sum_{\sigma}f_{\sigma}^{\dagger}f_{\sigma}=1
\end{eqnarray*}
\noindent where the bar on top of the index $\alpha$ indicates that it
transforms according to the complex conjugate representation. We used
\textit{slave-boson} language and  denoted
$\varepsilon_{q}\equiv E_{\Gamma_{3}}$ and $\varepsilon_{s}\equiv
E_{\Gamma_{6}}$ (let us also define
$\varepsilon\equiv\varepsilon_{s}-\varepsilon_{q}$ and $\Delta\equiv
V^2/2$).  Notice that the model has full $SU\!\left( 2\right) \otimes
SU\!\left( 2\right)$ internal symmetry. In the parameter regimes when
$|\varepsilon|\gg\Delta$, we can map it via a Schrieffer-Wolff
transformation into a quadrupolar two-channel Kondo model (with a
localized quadrupolar moment in the case when $\varepsilon
_{q}\ll\varepsilon_{s}$), or a magnetic two-channel Kondo model (with
a localized magnetic moment in the case when
$\varepsilon_{q}\gg\varepsilon_{s}$), in both cases with a coupling
constant given by $J=V^{2}/\left\vert
\varepsilon_{s}-\varepsilon_{q}\right\vert$ (assuming zero chemical
potential).


We will also consider a generalized version of the Hamiltonian,
the {\it multi-channel Anderson impurity model}.
In {\it slave-boson} language the Hamiltonian looks exactly as in the
two-channel case, but this time the indices take values according to:
(\textbf{i}) $\sigma=1,\ldots,N$ and (\textbf{ii})
$\alpha=1,\ldots,M$. If we follow the standard nomenclature we would
call $\sigma$ the spin index and $\alpha$ the channel index, and the
model will be the $SU\!\left( N\right) \otimes SU\!\left(M\right)$
version of the multi-channel Anderson model \cite{Cox&Z}.

As  in the two-channel case we can perform Schrieffer-Wolff
transformations to derive effective models in any of the two local moment
regimes of the problem. When $\varepsilon_{q}\ll\varepsilon_{s}$ (assuming
always zero electronic chemical potential) we obtain an `$N$-channel
$SU\!\left(M\right)$ Coqblin-Schrieffer model', whereas when
$\varepsilon_{s}\ll\varepsilon_{q}$ we obtain an `$M$-channel $SU\!\left(
N\right)$ Coqblin-Schrieffer model'.


This model has been extensively studied over the past fifteen years by
a variety of methods: Large-N methods \cite{cr93}, NCA \cite{sac98},
conserving T-matrix approximation \cite{kw98}, Monte Carlo \cite{sac98}
and Numerical Renormalization Group \cite{kc99} among others. All these
techniques have difficulties accessing the mixed valence regime.

We showed that this model is integrable and carried out a full
analysis by means of the Bethe Ansatz technique. That allowed us to
obtain exact results characterizing the ground state of the model as a
function of its parameters and very accurate numerical results for the
thermodynamics of the model for all temperatures and fields, as well
as all valence regimes, including the mixed valence. We stress that is
in the mixed valence regime where the Bethe Ansatz solution is most
crucial, since all the other techniques have difficulties accessing
it.

\section{The Thermodynamics of the Model}

In this section we present the results of the numerical solution of
the Thermodynamic Bethe Ansatz equations for the two-channel case
(i.e.  $N=M=2$). This is the first model
proposed in the context of \textrm{UBe}$_{13}$\cite{cox87}. We will
study the physics of the model on its own right and also briefly
discuss its  virtues and inadequacies to describe the physics of
the uranium compound that motivated it. The conclusion that we will
reach is that one needs to go beyond the simplest scenario given
by the two-channel model.

\bigskip

\begin{figure}[tbh]
\label{SvsT_Qzero}
\begin{center}
\includegraphics[height=2.5507in,width=4.1443in]{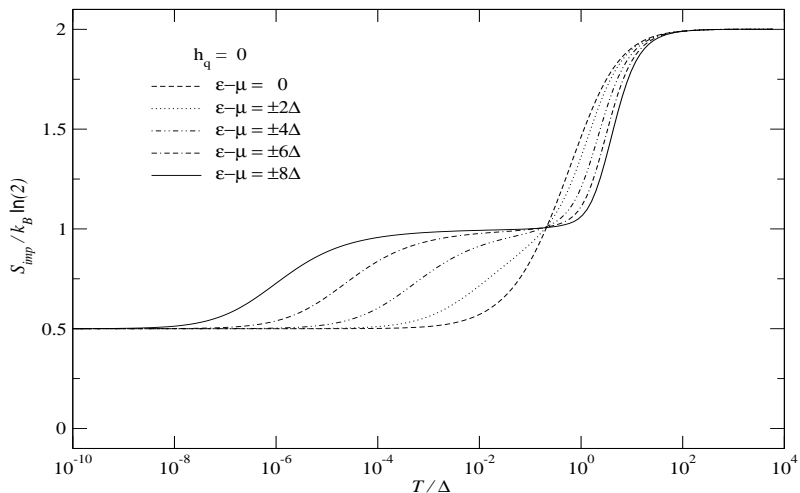}
\caption[Impurity contribution to the entropy at zero
field.]{Impurity contribution to the entropy at zero
field and as a function of temperature for different values of
$\varepsilon-\mu$. As the temperature goes to zero all curves approach
the universal value $k_{B}\ln\sqrt{2}$. Positive and negative values
of $\varepsilon-\mu$ fall on top of each other.}
\end{center}
\end{figure}

\begin{figure}[tbh]
\label{SvsT_Qem4}
\begin{center}
\includegraphics[height=2.5507in,width=4.1443in]{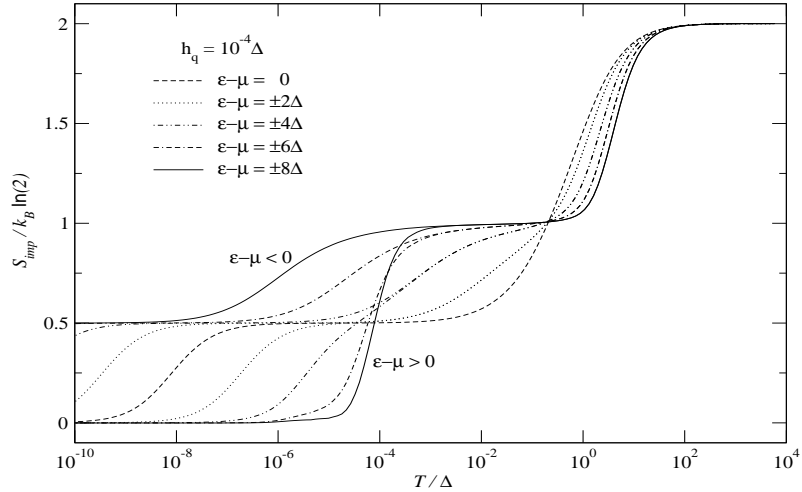}
\caption[Impurity contribution to the entropy at zero
field.]{Impurity contribution to the entropy in the
presence of an applied quadrupolar field and as a function of
temperature for different values of $\varepsilon-\mu$. As the
temperature goes to zero all curves approach zero. Positive and
negative values of $\varepsilon-\mu$ fall on top of each other only at
high temperatures and, asymptotically, for low temperatures.}
\end{center}
\end{figure}

In Fig.~\ref{SvsT_Qzero} we show the behavior of the impurity entropy
as a function of temperature for different values of
$\left(\varepsilon-\mu\right)$, and in Fig.~\ref{SvsT_Qem4} the reader
can see the effect of switching on an external field.


\begin{figure}[tbh]
\label{NcPlot}
\begin{center}
\includegraphics[height=2.5498in,width=3.8783in]{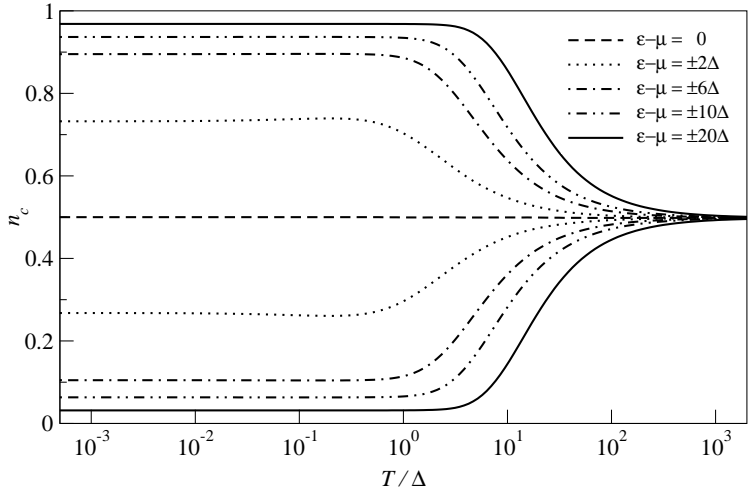}
\caption[Charge content of the impurity versus temperature for zero
field.]{Charge content of the impurity site as a
function of temperature for different values of $\varepsilon-\mu$. At
high temperatures all curves start from $n_{c}=1/2$ and flow, as the
temperature is lowered to specific values $n_{c}^{0}\left(
\varepsilon-\mu\right)$.}
\end{center}
\end{figure}

At high temperatures the entropy is
$S_{\mathrm{imp}}=k_{\mathrm{B}}\ln4$ in agreement with the size of
the impurity Hilbert space. For $|\varepsilon-\mu|\gg\Delta$ the impurity
entropy is \textit{quenched} in two stages. The degrees of freedom
corresponding to the higher energy doublet are frozen first. The
entropy becomes $S_{\mathrm{imp}}=k_{\mathrm{B}}\ln2$ and the system
is in a localized magnetic or quadrupolar moment regime depending on
the sign of $\left(\varepsilon-\mu\right)$. As the temperature is further
decreased, the remaining degrees of freedom undergo frustrated
screening leading to entropy
$S_{\mathrm{imp}}=k_{\mathrm{B}}\ln\sqrt{2}$. On the other hand, for
values of $|\varepsilon-\mu|\ll\Delta$, the quenching process takes place
in a single stage. The initial stage of the quenching process happens
when the system makes the transition from a high-energy state of
valence $n_{c}=\frac{1}{2}$ (i.e. when both states of valence are
equally likely) to the state of valence that will be present at
low-energies, $n_{c}^{0}(\varepsilon-\mu)$. This is further illustrated in
Fig.~\ref{NcPlot}.

As $\left(\varepsilon-\mu\right)$ is varied, the behavior interpolates
continuously between the magnetic ($n_{c}^{0}=1$) and the quadrupolar
($n_{c}^{0}=0$) scenarios. The zero temperature entropy is found to be
independent of $\varepsilon$ in accordance with our analytic
results \cite{ba02}. External fields, either magnetic or quadrupolar,
constitute relevant perturbations that drive the system to a Fermi
Liquid fixed point with zero entropy.

\begin{figure}[tbh]
\label{CvPlot}
\begin{center}
\includegraphics[height=2.5498in,width=3.9803in]{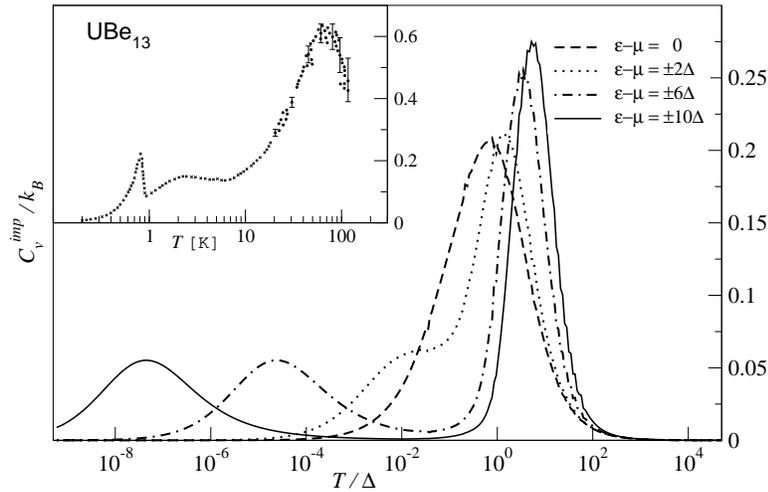}
\caption[Specific heat versus temperature at zero
field.]{Main plot: impurity contribution to the specific
heat as a function of temperature for different values of
$\varepsilon-\mu$. As this parameter approaches zero the Kondo
contribution (left) and the Schottky anomaly (right) collapse in a
single peak. Inset: experimental data for the $5f$-derived specific
heat of UBe$_{13}$.}
\end{center}
\end{figure}

In Fig.~\ref{CvPlot} we show the impurity contribution to the specific heat.
The two stage quenching process gives rise to two distinct peaks. The lower
temperature peak is the Kondo contribution centered around a temperature
$T_{l}\left( \varepsilon-\mu\right) $, whereas the higher temperature peak
--~often referred to as the Schottky anomaly~-- is centered around a
temperature $T_{h}\left( \varepsilon-\mu\right) $. Approximate expressions
for $T_{h,l}$ can be read off from the curves:
\[
T_{h,l}(\varepsilon)\approx\frac{\Delta}{a\pi^{2}}\ln(1+2a\,e^{\pm\frac{\pi
}{2\Delta}|\varepsilon-\mu|})
\]
with $1<a<4$. For large $|\varepsilon-\mu|$ the two peaks are clearly
separated and the \textit{area}\footnote{Here by \textit{area} we mean the
following integral: $\int C_{v}\left(  T\right)  ~d\left(  \ln T\right)  $. In
other words, the area as plotted, i.e. in a logarithmic scale.} under the
Kondo peak is $k_{B}\ln\sqrt{2}$ while that under the Schottky peak is
$k_{B}\ln2$.

\bigskip

As mentioned earlier, the model was proposed as a description for the
uranium ion physics of \rm{UBe}$_{13}$. It is expected to describe the
lattice above some coherence temperature. We provide in the inset of
Fig.~\ref{CvPlot} the experimental data for the $5\mathrm{f}$-derived
specific heat of the compound. It is obtained by subtracting from its
total specific heat, the specific heat of the isostructural compound
\rm{ThBe}$_{13}$ containing no $5\mathrm{f}$ electrons
\cite{felten86,felten87}. This way one is subtracting the phonon
contribution as well as the electronic contribution from electrons in
$\mathrm{s}$, $\mathrm{p}$ and $d$ shells (the procedure is quite
involved and we refer the reader to the cited articles for full
details). The sharp feature at $\sim0.8K$ signals the superconducting
transition of UBe$_{13}$ and falls outside the range where this
compound might be described by considering a single impurity model.

In order to be certain of having eliminated lattice effects it would
be better to carry out the measurements on \rm{U}$_{1-x}$\rm{Th}$_{x}%
$\rm{Be}$_{13}$. For $x>0.1$ the compound has no longer a
superconducting transition and the lattice coherence effects are
largely suppressed. Further, there are several experimental
indications that support the idea of an impurity model description of
the thoriated compound for a wide range of temperatures
\cite{aliev96}.

Concentrating on the temperature range containing the Kondo and
Schottky peaks we conclude that no values of $\varepsilon$ and
$\Delta$ of the 2-channel model yield a good fit.  Further, the
entropy obtained by integrating the weight under the experimental
curve falls between $k_{\rm{B}}\ln4$ and $k_{\rm{B}}\ln6$. This
suggests that a full description of the impurity may involve another
high energy multiplet (possibly a $\Gamma_{4}$ triplet, cf.~with the
work of Koga and Cox \cite{kc99}) close to the $\Gamma_{3}$ to yield
the peak for the Schottky anomaly, with the $\Gamma_{6}$ doublet
falling outside the range of measurements. The nature of the multiplet
could be deduced from further specific heat measurements. For an
n-plet degenerate with the $\Gamma_{3}$, one has an $SU\!(2)\otimes
SU\!(n+2)$ Anderson model and the area under $C_{\rm{v}}^{\rm{imp}}/T$
is then given by $k_{\rm{B}}\ln[\frac{n+4}{2}\sec\frac{\pi}{n+4}]$,
while if the n-plet is slightly split off the doublet, the area is
given by $k_{\rm{B}}\ln[\frac{n+4}{\sqrt{2}}]$ \cite{jaz98}. Also the
scale of energy involved is then determined by crystal field splitting
and corresponds to the scale observed. The split off non-magnetic
triplet (i.e. $n=3$ in the $f^2$ configuration seems, indeed, to be
the most reasonable scenario for $U Be_{13}$. We are currently working
in this direction.


\begin{acknowledgements}
We thank the organizers for the invitation to participate in the NATO
Advanced Research Workshop on Concepts in Electron Correlation.
\end{acknowledgements}

\bibliographystyle{prsty}
\bibliography{Proc_UBe13}

\end{article}
\end{document}